\documentclass[superscriptaddress,aps,pra,twocolumn]{revtex4}
\usepackage{bm}
\usepackage{ulem}
\usepackage{epsfig}
\usepackage{graphicx}
\usepackage{amssymb,amsmath,amsbsy,amsgen,amsfonts}    
\usepackage{dcolumn}
\usepackage{amsthm}
\usepackage{mathrsfs}
\usepackage{latexsym}
\usepackage{array}
\usepackage{color}
\usepackage{amstext}
\allowdisplaybreaks[1]
\usepackage{txfonts}
\usepackage{pifont}

\usepackage{epstopdf} 

\newcommand{\abs}[1]{|#1|}

\newcommand{\bra}[1]{\left\langle{#1}\right\vert}
\newcommand{\ket}[1]{\left\vert{#1}\right\rangle}

\newcommand{\be}{\begin{equation}}
\newcommand{\ee}{\end{equation}}
\newcommand{\ba}{\begin{array}}
\newcommand{\ea}{\end{array}}
\newcommand{\bqa}{\begin{eqnarray}}
\newcommand{\eqa}{\end{eqnarray}}

\DeclareSymbolFont{symbols}{OMS}{cmsy}{m}{n}

\begin{document}
\title{Quantum Plasmonic Sensing: Beyond the Shot-Noise and Diffraction Limit}
\author{Changhyoup Lee}
\email{changdolli@gmail.com}
\affiliation{Department of Physics, Hanyang University, Seoul, 133-791, South Korea}
\affiliation{Institute of Theoretical Solid State Physics, Karlsruhe Institute of Technology, 76131 Karlsruhe, Germany}

\author{Frederik Dieleman} 
\affiliation{Department of Physics, Imperial College London, London SW7 2AZ, UK}

\author{Jinhyoung Lee}
\affiliation{Department of Physics, Hanyang University, Seoul, 133-791, South Korea}

\author{Carsten Rockstuhl}
\affiliation{Institute of Theoretical Solid State Physics, Karlsruhe Institute of Technology, 76131 Karlsruhe, Germany}
\affiliation{Institute of Nanotechnology, Karlsruhe Institute of Technology, Karlsruhe, Germany}

\author{Stefan A. Maier}
\affiliation{Department of Physics, Imperial College London, London SW7 2AZ, UK}

\author{Mark Tame}
\email{markstame@gmail.com}
\affiliation{University of KwaZulu-Natal, School of Chemistry and Physics, Durban 4001, South Africa}
\affiliation{National Institute for Theoretical Physics (NITheP), KwaZulu-Natal, South Africa}
\date{\today}

\begin{abstract}
Photonic sensors have many applications in a range of physical settings, from measuring mechanical pressure in manufacturing to detecting protein concentration in biomedical samples. A variety of sensing approaches exist, and plasmonic systems in particular have received much attention due to their ability to confine light below the diffraction limit, greatly enhancing sensitivity. Recently, quantum techniques have been identified that can outperform classical sensing methods and achieve sensitivity below the so-called shot-noise limit. Despite this significant potential, the use of definite photon number states in lossy plasmonic systems for further improving sensing capabilities is not well studied. Here, we investigate the sensing performance of a plasmonic interferometer that simultaneously exploits the quantum nature of light and its electromagnetic field confinement. We show that, despite the presence of loss, specialised quantum resources can provide improved sensitivity and resolution beyond the shot-noise limit within a compact plasmonic device operating below the diffraction limit.
\end{abstract}


\maketitle
{\it Introduction---.}
Plasmonic excitations have attracted enormous interest in recent years from a variety of scientific fields due to their intriguing light-matter features and wide range of applications~\cite{Zia06,Ozbay06}. Plasmonic biosensing, in particular, is one of the most successful applications, with devices that outperform conventional ones that rely on ordinary photonic components~\cite{Homola99a,Lal07,Anker08}. Due to their high sensitivity, multiple surface plasmon resonance (SPR) sensing devices have been developed over the decades~\cite{Rothenhausler88,Homola99b,Dostalek05,Sepulveda06,Svedendahl09,Mayer11,Jorgenson93,Leung07}. The higher sensitivity of SPR sensors is achieved via a strong electromagnetic (EM) field enhancement at a metal surface, where its interaction with free electrons forms a surface plasmon that confines the field to a spatial domain below the diffraction limit~\cite{Raether88}. Such confinement is not possible with ordinary dielectric media~\cite{Takahara97}. Despite their practical realization and successful commercialization, the high sensitivity and associated resolution of SPR sensing are fundamentally limited by the discretized nature of light known as the shot-noise limit (SNL)~\cite{Caves81}. However, recently it has been shown that the SNL can be beaten by using quantum states of light having a super- or sub-Poissonian photon distribution, or intermode entanglement~\cite{Sahota15}, and an appropriate type of measurement -- a strategy known as quantum metrology~\cite{Giovannetti04}. A number of impressive experiments have already demonstrated the basic working features of quantum metrology using multiphoton states in bulk optics~\cite{Kuzmich98,Fonseca99,Bouwmeester04,Mitchell04,Walther04}, integrated optics~\cite{Matthews09} and sensing biological systems~\cite{Crespi12,Taylor13}. A question naturally arises about whether such quantum techniques could be employed in plasmonic sensors in order to further enhance their capabilities. Here, absorption constitutes a significant challenge that usually causes a degradation of the quality of a quantum resource~\cite{Tame13}.

\begin{figure}[t]
\centering
\includegraphics[width=8cm]{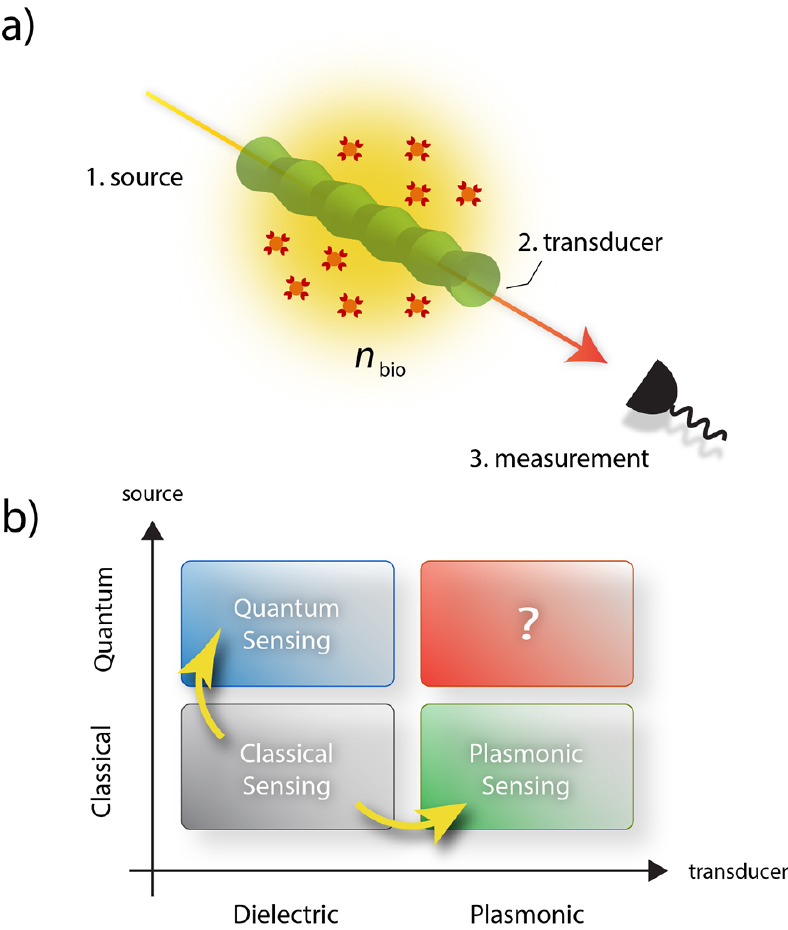}
\caption{General scenario for photonic sensing. a) For a properly chosen signal, one measures and analyzes the output light from a transducer. An example of a biosensor is given, where the transducer encodes onto the light signal changes in the biological medium. b) Four regions in which photonic sensing devices operate, distinguishing the use of quantum or classical techniques in the signal and measurement parts, and the use of dielectric or plasmonic material in the transducer part. The enhancement of sensitivity has been known to be achieved through the yellow arrows, whereas there is an intriguing region in the top right that has recently begun to attract attention, called quantum plasmonic sensing~\cite{Kalashnikov14, Fan15, Pooser16, Taylor16}.
}
\label{fig1} 
\end{figure}

Very recently work has shown the possibility of reducing quantum noise in plasmonic sensing by using a two-mode quadrature squeezed state in a prism configuration~\cite{Fan15,Pooser16} and in a nanoparticle array~\cite{Kalashnikov14}. However, the role of quantum effects in more general plasmonic sensing devices at the few-photon level is not well understood. To address this, we begin with a concept of quantum plasmonic sensing that utilizes both quantum features of resource states at the few-photon level and the EM field enhancement offered by plasmonic structures. We show how the combination of quantum and plasmonic aspects enables one to improve the sensitivity of a device beyond the SNL, while keeping its compactness on scales below the diffraction limit. We highlight the feasibility of our approach by examining the minimum resolution of parameter estimation in an example interferometer-based plasmonic biosensor. Here, we consider waveguides that have numerous attractive features geared toward the design of compact, mobile, broadband and integratable biosensors. Our analysis shows the beneficial role that quantum effects can play in a plasmonic sensor, despite the presence of loss. The techniques developed can be applied to many other plasmonic sensing platforms and thus we expect this work to stimulate a variety of further investigations beyond conventional quantum metrology and classical plasmonic sensing~\cite{Taylor16}.

{\it The concept of quantum plasmonic sensing---.}
We begin with the general scenario for photonic sensing shown in Fig.~\ref{fig1}a, which is divided into three stages: (i) a signal preparation where an incident light field is prepared, (ii) a transducer that encodes the information on the parameter to be measured onto the output signal, and (iii) a measurement that analyzes the output signal from the transducer. A biological setting is chosen as an example, where a physicochemical transducer encodes the information of surrounding biological objects onto the output signal. For other settings the transducer may take a different form, such as for mechanical~\cite{Carrascosa06, Hoff13, Pooser15}, electrical~\cite{Dolde11}, or magnetic parameters~\cite{Taylor08,Maze08}. In the classical measurement scenario, a classical source is used for the input signal, a dielectric medium represents the transducer and a classical intensity measurement is performed. An enhancement of sensitivity can be obtained here via two directions: First, plasmonic effects can be employed in the transducer by using a metallic medium providing a strong EM field enhancement. This enables a much higher sensitivity compared to the field in a conventional dielectric medium, as a change of environment produces a larger change of the mode properties of surface plasmons compared to photons~\cite{Homola99a,Lal07,Anker08}. Second, the signal and measurement parts can be replaced by quantum elements. For example, it has been shown that states known as NOON states~\cite{Boto00} or quadrature-squeezed states~\cite{Caves81} can improve the minimum resolution of parameter estimation beyond the SNL by using an appropriate measurement scheme. Such quantum strategies have been employed for biosensing recently to minimize the shot-noise associated with the random arrival of photons at a detector~\cite{Taylor13,McGuinness12}. Even more recently, the use of plasmonic elements has begun to be considered with the above quantum strategies, showing the capability of beating the SNL~\cite{Kalashnikov14, Fan15, Pooser16}. However, it is not entirely clear how quantum techniques can be incorporated into plasmonic sensing for further improving the sensitivity with finite photon number states, even though it has been demonstrated that properties such as quantum coherence can be preserved in plasmonic systems~\cite{Altewlscher02,Moreno04,Huck09,Lawrie13}. It is nontrivial that sensitivity beyond the SNL is achievable in such a lossy, open quantum system. As we will show, quantum plasmonic sensing is complementary to both classical plasmonic and quantum techniques for improving the sensitivity of photonic sensors, as depicted in Fig.~\ref{fig1}b. However, the key merit is that it improves functionality by beating the SNL in a subdiffraction scale system. 

\begin{figure*}[t]
\centering
\includegraphics[width=16cm]{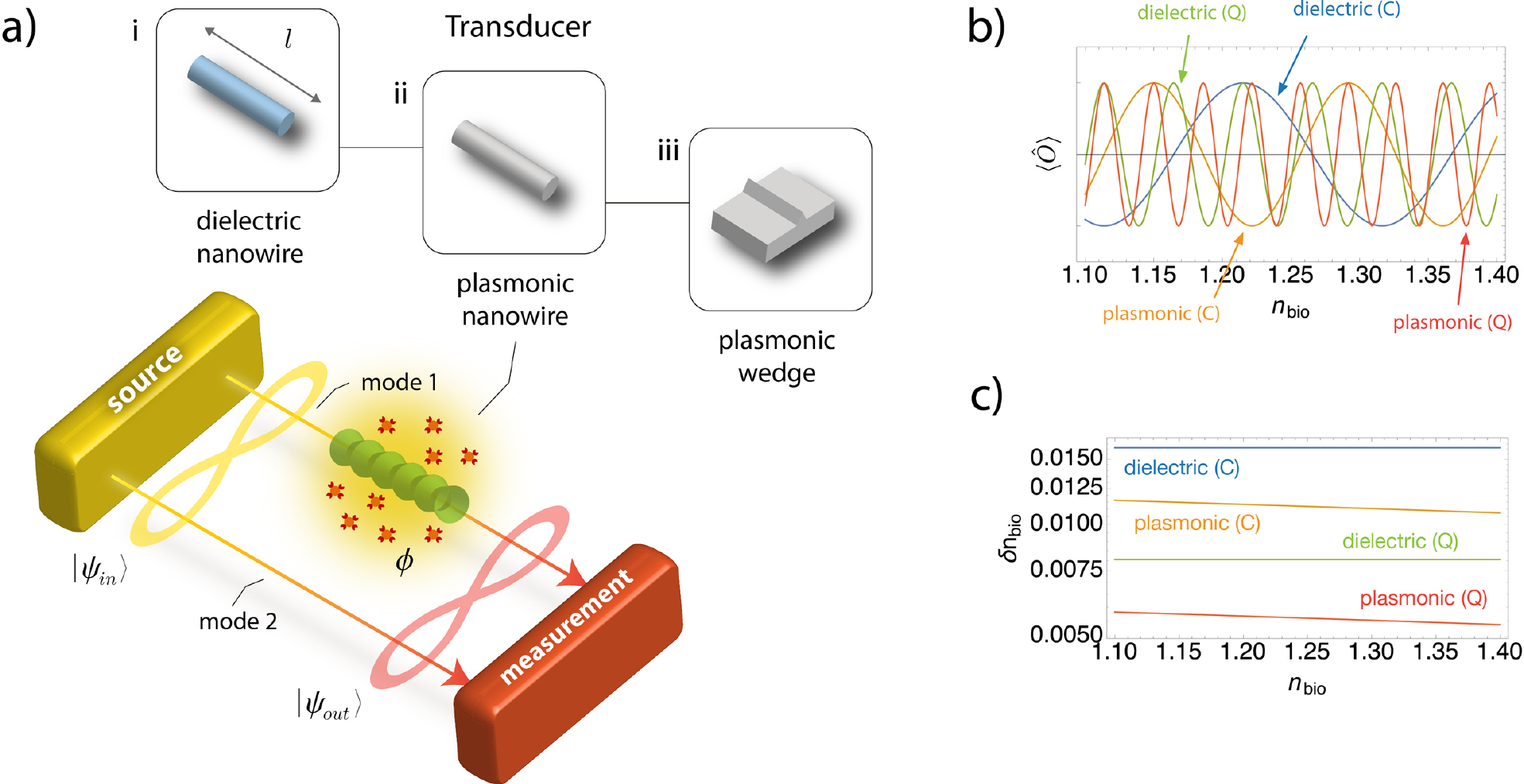}
\caption{Quantum plasmonic sensing. a) General two-mode interferometer with one arm in a nanowire waveguide. A quantum or classical state from a source stage is fed into the interferometer. The sensing arm (mode 1) is embedded in an environment and the signal acquires a phase change $\Delta \phi$ due to changes in the refractive index, $n_{\rm bio}$, during its propagation, modifying the output signal at the measurement stage. b) The expectation values of an observable $\langle \hat{O} \rangle$, where $\hat{O}=\hat{M}$ (with $\ket{\psi_{\rm out}}_{\rm classical}$) and $\hat{A}$ (with $\ket{\psi_{\rm out}}_{\rm quantum}$), optimized for classical (C) and quantum (Q) sensors, respectively. Here, an average photon number of $N=4$ is used to show that the quantum plasmonic case (the red curve) oscillates more rapidly than all others, implying that a small change of $n_{\rm bio}$ induces a large detectable change in the monitored output signal. In this example, we consider both dielectric and lossless silver metallic nanowires, with a core radius of $50~{\rm nm}$ and a length of $l=4~{\rm \mu m}$ at $\lambda_{0}=810~{\rm nm}$, where $n_{\rm core}=1.4475$ and $n_{\rm core}(\omega)=\sqrt{\epsilon_{m}(\omega)}$ from the experimental data in Ref.~\cite{Rakic98}. c) The minimum resolution, $\delta n_{\rm bio}$, shows that quantum plasmonic sensing exhibits the best performance.
}
\label{fig2} 
\end{figure*}

{\it Simulation---.} 
We illustrate the basic concept of quantum plasmonic sensing in Fig.~\ref{fig2}a, where a two-arm interferometer -- one of the most successful photonic sensing techniques~\cite{Lin97,Sepulveda06,Hoa07} -- is employed with a nanowire waveguide in one arm. We focus on a nanowire structure initially as it is a well studied geometry with many applications in plasmonic circuitry~\cite{Gramotnev10,Bozhevolnyi06} and has a high level of miniaturization for sensing with an accessible interrogation area~\cite{Homola99a}. The interferometric sensor consists of source and measurement parts and a transducer part for one arm. The transducer consists of a dielectric or metallic nanowire with refractive index $n_{d}=1.4475$ (doped silica) or $n_{m}(\omega)=\sqrt{\epsilon_{m}(\omega)}$ (silver) given by experimental data~\cite{Rakic98}, respectively. The nanowire is surrounded by a biological medium with refractive index $n_{\rm bio}$, whose value varies due to changes in the concentration of a biological analyte. We choose a range for $n_{\rm bio}$ that ensures single-mode operation in the waveguides (see Appendix A). The change in the biological medium changes the wavenumber $k$ of the waveguide mode, and the change of the wavenumber changes the relative accumulated phase of the fields, $\phi$, between both arms, which can be measured in the output signal via an interference measurement. Here, the wavenumber $k$ for dielectric and metallic nanowires is determined by a characteristic equation~\cite{Takahara97,Stockman04,Saleh07} (see Appendix A). From the measurement, we aim to estimate the refractive index unit (RIU) $n_{\rm bio}$ with the smallest detectable refractive index change $\delta n_{\rm bio}$, assuming for simplicity that there is no scattering when the input signal enters into the sensing region in the first arm.

A reference point for classical sensing can be found by considering the entire device as a Mach-Zehnder (MZ) interferometer. Here, a coherent state $\ket{\alpha}$, written in the Fock-state basis $\ket{m}$ as 
\begin{equation}
\ket{\alpha}={\rm e}^{-\abs{\alpha}^{2}/2} \sum_{m=0}^{\infty} \frac{\alpha^{m}}{\sqrt{m!}}\ket{m}
\end{equation} 
with mean photon number $\abs{\alpha}^{2}=N$, is fed into one input port of the first beamsplitter of the MZ and a vacuum state fed into the other. The output of this beamsplitter constitutes the source stage. An intensity-difference measurement $\hat{M}=\hat{I}_{1}-\hat{I}_{2}$ is performed by using the second beamsplitter of the MZ placed after the sensing region and measuring the output intensities. This constitutes the measurement stage. The above classical sensing strategy is optimal in that it leads to the SNL on the resolution $\delta n_{\rm bio}$.

On the other hand, by using quantum techniques one can consider a NOON state~\cite{Mitchell04,Walther04,Afek10} generated at the source stage, {\it i.e.} 
\begin{equation}
\ket{\psi_{\rm in}}=\frac{1}{\sqrt{2}}\left(\ket{N0}_{12}+\ket{0N}_{12}\right),
\end{equation}
where $N$ denotes the number of photons. The observable $\hat{A}=\ket{0,N}\bra{N,0}+\ket{N,0}\bra{0,N}$ can be used for the quantum measurement, which together with the NOON state allows one to reach the Heisenberg limit (HL) for $\delta n_{\rm bio}$ in the absence of photon loss~\cite{Giovannetti04,Mitchell04}.

In Fig.~\ref{fig2}b, we present the measurement signals $\langle \hat{M} \rangle=M_{0}{\rm cos}(\phi(n_{\rm bio}))$ and $\langle \hat{A} \rangle=A_{0}{\rm cos}(N\phi(n_{\rm bio}))$ simulated for the classical scenario using a coherent state and the quantum scenario using a NOON state, respectively. The state $\ket{\psi_{\rm out}}$ generated from encoding $\phi$ onto the input state $\ket{\psi_{\rm in}}$ is used to calculate $\langle ... \rangle$, {\it i.e.} $\ket{\psi_{\rm out}}_{\rm classical}=\ket{ \frac{1}{2} \alpha({\rm e}^{i\phi} -1)}_{1}\ket{\frac{1}{2} i \alpha({\rm e}^{i\phi} +1)}_{2}$, and $\ket{\psi_{\rm out}}_{\rm quantum}=\frac{1}{\sqrt{2}}(e^{iN\phi}\ket{N0}_{12}+\ket{0N}_{12})$  (see Appendix B). Here, $N=4$ has been chosen and $\phi(n_{\rm bio})$ denotes the relative phase accumulated during propagation (free-space wavelength $\lambda_{0}=810~{\rm nm}$ chosen as an example) along a dielectric and silver nanowire with a core radius of $50~\rm{nm}$ and length $l=4~{\rm \mu m}$. The lateral confinement of the field of the dielectric nanowire is diffraction limited ($\sim \lambda_0/n_d$), whereas that of the metallic nanowire is not ($\ll \lambda_0$)~\cite{Takahara97}. For the relative phase picked up, we have $\phi(n_{\rm bio})=\beta(n_{\rm bio}) \times l$, where the propagation constant $\beta(n_{\rm bio})\equiv {\rm Re}[k]$ is a function of $n_{\rm bio}$ (see Appendix A). Here, we have considered a lossless silver nanowire, {\it i.e.} ${\rm Im}[k]=0$. We consider the impact of losses later. The main purpose at this stage is to show the difference between classical and quantum techniques, and the use of dielectric and plasmonic systems. It can be seen in Fig.~\ref{fig2}b that the expectation value for the quantum plasmonic case oscillates far more rapidly than the others, implying that a small change of $n_{\rm bio}$ induces a large detectable change of the measurement signal. It may seem like one can resolve an infinitesimal change of $n_{\rm bio}$ by simply measuring the change of a given measurement signal, but this is not the case as the curves in Fig.~\ref{fig2}b become naturally blurred when quantum fluctuations are involved. Therefore, in Fig.~\ref{fig2}c we evaluate the minimum resolution of the refractive index change achievable from an observable $\hat{O}$ ($=\hat{A}$ or $\hat{M}$ for quantum or classical scenarios) with quantum fluctuations included. The resolution is obtained by the linear error propagation method~\cite{Durkin07} as
\begin{eqnarray}
\delta n_{\rm bio} = \frac{\Delta \hat{O}}{ \abs{\partial \langle \hat{O} \rangle /\partial n_{\rm bio}}},
\label{resolution}
\end{eqnarray}
where $\Delta \hat{O}=(\langle \hat{O}^{2} \rangle - \langle \hat{O}\rangle^{2})^{1/2}$. Here, the parameter of interest in our biosensing scenario is $n_{\rm bio}$ instead of the relative phase $\phi$ -- the usual quantity considered in quantum metrology~\cite{Giovannetti04}. Its corresponding resolution $\delta n_{\rm bio}$ depends on the waveguide material. The behaviours seen in Fig.~\ref{fig2}c clearly show that for the quantum plasmonic case the resolution $\delta n_{\rm bio}$ is smallest compared to the others. This implies that quantum plasmonic sensing can outperform both standard dielectric quantum metrology and classical plasmonic sensing within the same parameter regime. We note that the quantum case yields a material-dependent HL, $\delta n_{\rm bio}^{\rm (HL)} = \frac{1}{N}\abs{\frac{\partial \phi}{\partial n_{\rm bio}}}^{-1}$, which has a factor $\sqrt{N}$ improvement over the classical case with a material-dependent SNL, $\delta n_{\rm bio}^{\rm (SNL)} = \frac{1}{\sqrt{N}}\abs{\frac{\partial \phi}{\partial n_{\rm bio}}}^{-1}$, the origins of which we discuss in detail below.

\begin{figure}[t]
\centering
\includegraphics[width=8.5cm]{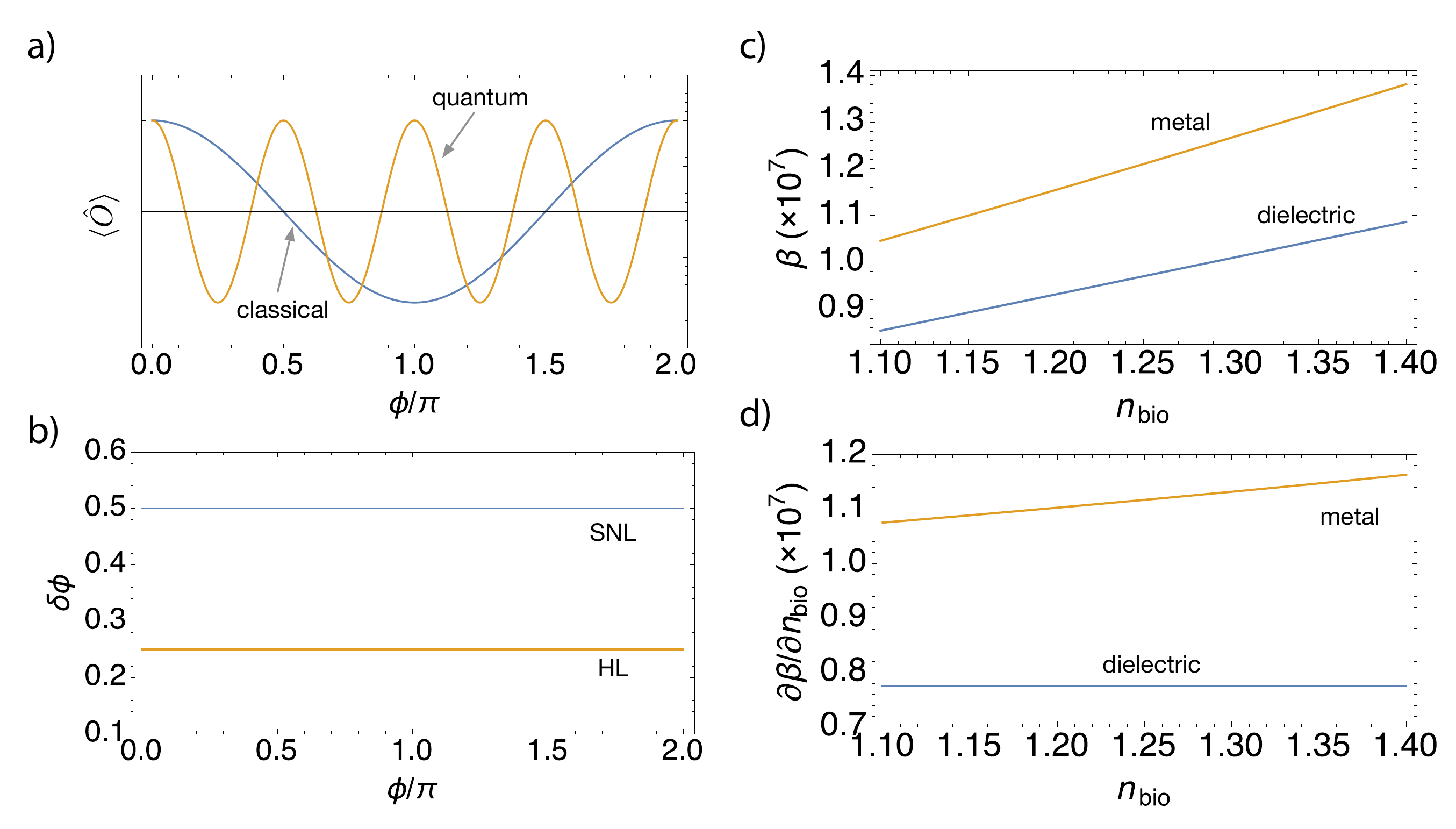}
\caption{ 
Roles of quantum and plasmonic effects. Quantum-enhanced sensitivity comes from the source and measurement stages, which are responsible for how sensitive the initial state and the observable are with respect to the phase. a) Comparison of the classical and quantum metrology scenarios in terms of the expectation values of $\hat{A}$ and $\hat{M}$. b) Minimum estimation precision, $\delta \phi$, corresponding to the measurements in panel a. In panel a, the expectation value of $\langle \hat{A} \rangle$ oscillates more rapidly than that of $\langle \hat{M} \rangle$; the classical case leads to the SNL and the quantum case leads to the HL in panel b, {\it i.e.} $\delta \phi^{({\rm SNL})}=1/\sqrt{N}$ and $\delta \phi^{({\rm HL})}=1/N$. The plasmonic enhanced sensitivity comes from the transducer, which can be seen in terms of the relation between the propagation constant $\beta$ and $n_{\rm bio}$. c) $\beta$ for the lossless plasmonic and dielectric waveguides. d) Slope of $\beta$ over $n_{\rm bio}$ showing the rate of change.}
\label{fig3} 
\end{figure}

{\it Quantum and plasmonic features combined for enhanced sensing---.}
We now look at how quantum resources enable plasmonic sensing to go beyond the SNL. The interferometric setup in Fig.~\ref{fig2}a has the ability to quantitatively detect a phase change $\Delta\phi$ induced by a change of the propagation constant, {\it i.e.} $\Delta\phi=\Delta\beta\times l$, where $\Delta\beta$ is induced by a variation in the analyte. Thus, the chosen material in the transducer is only responsible for how sensitively it accumulates $\Delta\phi$ as $n_{\rm bio}$ changes. On the other hand, the quantum source and measurement are responsible for how sensitively the chosen state and measurement stage respond to $\Delta\phi$. Such separate roles are manifested in the \textit{sensitivity}, defined as the ratio of the change in sensor output $\langle\hat{O}\rangle$ to the change in $n_{\rm bio}$, which can be written by the chain rule as
\begin{eqnarray}
{\cal S}=\frac{\partial \langle \hat{O} \rangle}{\partial n_{\rm bio}}=\frac{\partial \langle \hat{O} \rangle}{\partial \phi}\frac{ \partial \phi}{\partial n_{\rm bio}},
\label{sensitivity}
\end{eqnarray} 
where the expectation value $\langle\hat{O}\rangle$ is assumed to have only $\phi$-dependence with respect to $n_{\rm bio}$. The first term on the right hand side describes the sensitivity of the output $\langle\hat{O}\rangle$ to $\phi$, whereas the second term describes the sensitivity of $\phi$ to $n_{\rm bio}$. Consequently, Eq.~(\ref{resolution}) can be rewritten as 
\begin{eqnarray}
\delta n_{\rm bio}= \delta \phi ~\Big\vert \frac{\partial \phi}{\partial n_{\rm bio}}\Big\vert^{-1},
\label{finalresolution}
\end{eqnarray}
where $\delta\phi=\Delta\hat{O}/\abs{\partial\langle\hat{O}\rangle/\partial\phi}$ denotes the minimum resolution of the phase and does not depend on the waveguide material, provided that $\ket{\psi_{\rm out}}$ can be written as a function of $\phi$ only. Note that it is the nonclassical nature of the source and the measurement that decreases $\delta\phi$ below the SNL, which is clearly seen in Figs.~\ref{fig3}a and b, where we reproduce well known behaviours of $\langle\hat{O}\rangle$ and $\delta\phi$ for classical and quantum metrology with the same input states and measurements used in Fig.~\ref{fig2}. On the other hand, the sensitivity $\partial\phi/\partial n_{\rm bio}~(=l\times\partial\beta/\partial n_{\rm bio})$ depends on the material used and can be increased by a plasmonic transducer. In Figs.~\ref{fig3}c and d, we show $\beta$ and its slope change with increasing RIU for dielectric and metallic waveguides. The enhanced sensitivity of $\beta$ is a result of the strong field confinement for the plasmonic mode, making it more sensitive to $\Delta n_{\rm bio}$. In other words, the resolution $\delta\phi$ is improved by properties of a chosen input state and measurement, while the sensitivity $\partial\phi/\partial n_{\rm bio}$ is improved by the mode properties of the transducer. The combined effect of these quantum and plasmonic features is what leads to the results seen in Fig.~\ref{fig2}c.

Our analysis can be generalized to any kind of plasmonic setup for which the sensitivity and the minimum resolution can be rewritten in terms of a parameter $X$ as $\tilde{{\cal S}}=\frac{\partial\langle\hat{O}\rangle}{\partial X}\frac{\partial X}{\partial n_{\rm bio}}$ and $\delta\tilde{n}_{\rm bio}=\delta X\abs{{\frac{\partial X}{\partial n_{\rm bio}}}}^{-1}$, where $\delta X=\Delta\hat{O}/\abs{\partial\langle\hat{O}\rangle/\partial X}$ denotes the minimum resolution of parameter estimation. The enhancement of $\partial X/\partial n_{\rm bio}$ depends on the material, the modulation technique and the surface plasmon excitation method~\cite{Piliarik09}. On the other hand, $\partial\langle\hat{O}\rangle/\partial X$ and $\delta X$ depend on the quantum input $\ket{\psi_{\rm in}}$ and the measurement $\hat{O}$, where the output $\ket{\psi_{\rm out}}$ is generated from encoding $X$ onto the input $\ket{\psi_{\rm in}}$. For example, the widely used Kretschmann configuration could replace the transducer shown in Fig.~\ref{fig1}a~\cite{Fan15,Pooser16} and the reflection coefficient $\abs{R}^{2}$ used as the effective parameter $X$. In this case, the refractive index change would not be picked up as a phase, but rather as an intensity (or peak angular position). However, one could also consider embedding the Kretschmann configuration directly within an interferometer~\cite{Kashif14}, and the change picked up as a phase, bringing this method inline with the interferometric setting we have described. These more general expressions provide a better understanding of the specific roles that quantum and plasmonic features play, and enable the efficient optimization of quantum plasmonic sensing. In addition to enhanced sensitivity and resolution, there are other advantages of using plasmonics, {\it e.g.}~a small-sized mode volume below the diffraction limit that conventional photonics cannot achieve. This is important since a highly miniaturized sensor is commonly required to measure tiny organic molecules within a limited interaction area~\cite{Homola99a}. The combination of the reduced shot-noise of a quantum resource and the enhanced sensitivity provided by plasmonics guarantees that quantum plasmonic sensing can, in principle, go beyond both the shot-noise and the diffraction limit.

{\it Realistic scenario including loss---.}
We now show that quantum plasmonic sensing remains able to beat the SNL even when realistic metallic losses are included. To do this, we require an optimal quantum state for the source for a given amount of loss. The NOON state previously studied is extremely fragile to loss and is not an optimal quantum state, resulting in a much worse resolution than the SNL even for moderate loss~\cite{Dorner09}. Assuming the optimal measurement will be performed, we focus on optimizing the input state. In this case, the minimum resolution is given by the Cram\'er-Rao bound according to quantum parameter estimation theory~\cite{Braunstein94}, {\it i.e.} 
\begin{equation}
\delta\phi=F_{Q}^{-1/2},
\end{equation}
where the quantum Fisher information, $F_{Q}$, represents a measure of the amount of information that a state contains about $\phi$ with respect to the optimized measurement over all possible schemes (see Appendix C). We optimize the coefficients of an input state with fixed $N$ written as 
\begin{equation}
\ket{\psi_{\rm in}}=\sum_{n=0}^{N}c_{n}\ket{n,N-n},
\end{equation}
such that $F_{Q}$ is maximized and $\delta \phi$ is minimized~\cite{Dorner09}.

As possible plasmonic waveguides we consider the nanowire waveguide previously studied and a wedge waveguide~\cite{Pile05}, as shown in Fig.~\ref{fig2}a. Wedge waveguides have recently been shown to be highly beneficial for plasmonic devices in the quantum regime due to their high field confinement~\cite{Kress15} and broadband response over a wide operating range, a key requirement for a good biosensor, allowing one to avoid frequencies where the analyte is absorbing. The amount of loss in the waveguides is determined by ${\rm Im}[k]$ and $l$, with $l=4~{\rm \mu m}$ chosen as an example. We use a beamsplitter model for including loss, where a fictitious beam splitter with a transmitivity $\eta={\rm exp}(-2{\rm Im}[k]l)$ is inserted in one arm of the interferometer. Such a model is also valid for loss occurring during the phase acquisition in a metallic nanowire since the loss operation and the phase accumulation commute with each other~\cite{Dorner09}. The parameter $F_{Q}$ is then given as a function of the set $\{x_{n}=\abs{c_{n}^{2}}\}$ and $\eta$ (see Appendix C). For the nanowire, we consider the same range of $n_{\rm bio}$ as before in order to aid comparison of the results. On the other hand, for the refractive index near the wedge waveguide, in order to give a more realistic scenario, we consider $n_{bio}=n_{s}+A\times C$, where $n_{s}=1.333$ denotes water as a solvent, and $A=0.00182$ and $C$ represent Bovine Serum Albumin (BSA) as a solute~\cite{Leung07} and the number of grams of BSA solute per $100{\rm ml}$ of solution, respectively. For the wedge waveguide $C$ is varied from $0\%$ to $60\%$, yielding $n_{\rm bio}$ ranging from $1.333$ to $1.4422$. For the nanowire, $n_{\rm bio}$ is between $1.1$ and $1.4$ as before, ensuring that only a single mode exists for a radius of $50~{\rm nm}$. For the wedge waveguide, the top angle is $70.6^{\circ}$ and the bottom angles are $54.7^{\circ}$ (see Appendix D). The plasmon mode sits on top of the wedge and the height can be set arbitrarily small down to $\sim 50~{\rm nm}$ for $\lambda_0=810~{\rm nm}$, after which the mode has a significant presence at the bottom edges~\cite{Pile05,Kress15}.

\begin{figure}[t]
\centering
\includegraphics[width=8.75cm]{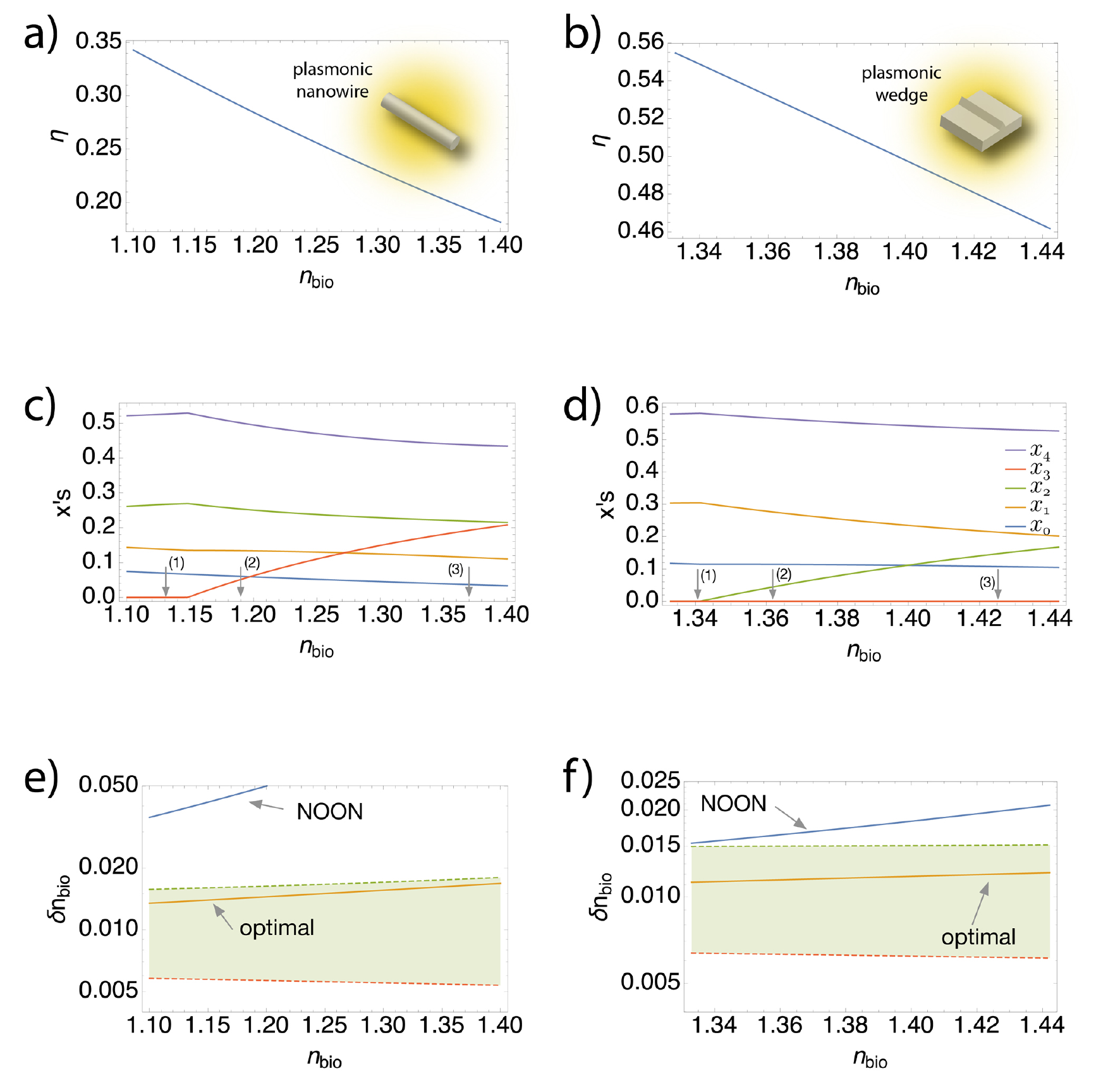}
\caption{ 
Realistic quantum plasmonic sensing. A lossy nanowire (left column) and a wedge waveguide (right column) are considered, where loss is modelled as a fictitious beamsplitter with a transmitivity $\eta={\rm exp}(-2{\rm Im}[k] l)$ for a propagation length $l$. a) For propagation with $l=4~{\rm \mu m}$, $\eta$ is obtained in terms of $n_{\rm bio}$ by solving the characteristic equation for the nanowire~\cite{Takahara97, Stockman04}. b) FEM simulation via COMSOL is used for the wedge waveguide (top angle $70.6^{\circ}$ and bottom angles $54.7^{\circ}$). c) The optimal set $\{ x_{n} \}$ for a state with a definite photon number $N=4$ is shown for the nanowire waveguide. d) The optimal set $\{ x_{n} \}$ for the wedge waveguide. e) The optimal resolution imposed by the Cram\'er-Rao bound for the NOON state and the optimized state for a given $\eta$ for the nanowire. f) The optimal resolution for the wedge waveguide. The shaded area in panels e and f is bounded at the top by the standard interferometric limit (SIL), corresponding to the SNL but optimized for an unbalanced beamsplitter to minimize the resolution in the presence of losses~\cite{Cooper11}. The area is bounded at the bottom by the HL. A line within the shaded area shows an improvement over classical plasmonic sensing.
}
\label{fig4} 
\end{figure}

For the respective ranges of $n_{\rm bio}$, we present the transmission coefficient $\eta$ in Figs.~\ref{fig4}a and b. For each $\eta$, depending on $n_{\rm bio}$, the optimal distributions of $\{x_{n}\}$ for $N=4$ are shown in Figs.~\ref{fig4}c and d, for which $F_{Q}$ is maximized, yielding the optimal minimum resolutions in Figs.~\ref{fig4}e and f. The optimal $x_{n}$ coefficients define the `optimal state' for sensing and are different depending on the amount of loss, but their relative phases are not important~\cite{Dorner09}. We also compare the optimal minimum resolutions with the HL (lower dashed lines) and the standard interferometric limit (SIL) (upper dashed lines), which corresponds to the SNL, but optimized using an unbalanced beamsplitter in order to minimize resolution in the presence of losses~\cite{Cooper11}. It can be clearly seen that the resolution with the NOON state is much worse than the SIL given as 
\begin{equation}
\delta n_{\rm bio}^{\rm (SIL)}=\frac{1+\sqrt{\eta}}{2\sqrt{N\eta}}\Big\vert\frac{\partial \phi}{\partial n_{\rm bio}}\Big\vert^{-1}, 
\end{equation}
whereas the optimal state beats the SIL in both plasmonic waveguides.

\begin{figure}[t]
\centering
\includegraphics[width=8.5cm]{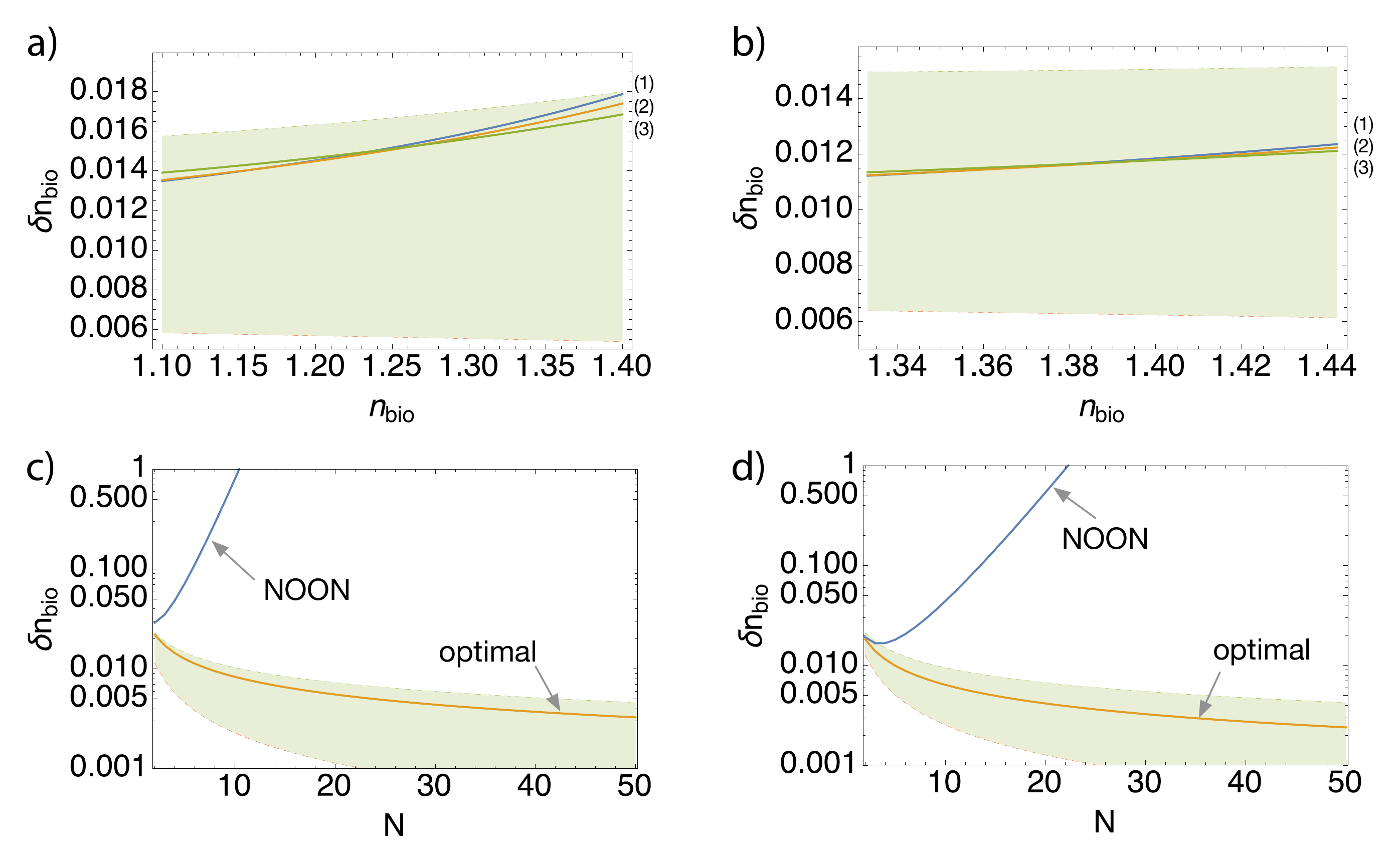}
\caption{Resolutions of optimal states and increasing average number of photons. Input states optimized for a certain amount of loss still beat the SIL over the whole range of $n_{\rm bio}$ for the lossy plasmonic nanowire and wedge waveguides considered in Fig.~\ref{fig4}. a) Three points are randomly chosen from Fig.~\ref{fig4}c): (1) $n_{\rm bio}=1.13$, (2) $n_{\rm bio}=1.19$ and (3) $n_{\rm bio}=1.37$. For these points, the optimal states found with the corresponding amount of loss are used as the input states and the resolutions are obtained over the whole range of $n_{\rm bio}$ for the nanowire waveguide. b) Same as panel a, but for the wedge waveguide. Here the points correspond to those in Fig.~\ref{fig4}d): (1) $n_{\rm bio}=1.34392$, (2) $n_{\rm bio}=1.36576$ and (3) $n_{\rm bio}=1.43128$. c) The resolution $\delta n_{\rm bio}$ with increasing $N$ for the NOON state, optimal state, SIL, and HL cases show that the optimal states remain beyond the SIL, regardless of $N$ for the nanowire. d) Same as panel c, but for the wedge waveguide. In both panels c and d, the value of $n_{\rm bio}$ corresponds to point (2) in Figs.~\ref{fig4}c and d, respectively. The quantum enhancement in both plots (c and d) is reduced with increasing $N$, as the gap between the optimal state resolution and the SIL decreases with $N$. A line within the shaded area shows an improvement over classical plasmonic sensing.
}
\label{fig5} 
\end{figure}

The optimal state is different depending on the amount of loss, so for experimental relevance it is important to check if a given input state optimized for a certain amount of loss still beats the SIL over the whole range of $n_{\rm bio}$ measured. Figs.~\ref{fig5}a and b present the results of three points chosen from Figs.~\ref{fig4}c and d, respectively, showing that the chosen states remain beyond the SIL over the respective ranges of $n_{\rm bio}$. 

It is worth noting that the quantum dielectric case where losses are absent or nearly negligible, shown in Fig.~\ref{fig2}c, provides smaller a resolution $\delta n_{\rm bio}$ than the case of quantum plasmonic sensing including loss. However, this does not mean that the quantum dielectric case is always the best strategy because it is diffraction limited and cannot be used for a sensing on scales far below the operating wavelength. In such a scenario, the use of quantum resources in a plasmonic system would be the best strategy, although at the cost of the sensing resolution.

We also investigate the resolution $\delta n_{\rm bio}$ as $N$ increases for the NOON state, the optimal state, the SIL and the HL. These more general results are shown in Figs.~\ref{fig5}c and d, where the optimal states remain beyond the SIL (upper dashed line), regardless of $N$. It should be noted that the quantum enhancement is reduced with increasing $N$, {\it i.e.}~in Figs.~\ref{fig5}c and d the gap between the HL and the SIL decreases with increasing $N$, $\delta n_{\rm bio}^{\rm (SIL)}-\delta n_{\rm bio}^{\rm (HL)}=\frac{1}{\sqrt{N}}(\frac{1+\sqrt{\eta}}{2\sqrt{\eta}}-\frac{1}{\sqrt{N}})\abs{\frac{\partial\phi}{\partial n_{\rm bio}}}^{-1}$, while the optimal state resolution remains at a roughly fixed distance below the SIL. On the other hand, the relative difference, defined by $(\delta n_{\rm bio}^{\rm (SIL)}-\delta n_{\rm bio}^{\rm (HL)})/\delta n_{\rm bio}^{\rm (SIL)}$, approaches unity in the limit of large $N$, while the other relative difference, defined by $(\delta n_{\rm bio}^{\rm (SIL)}-\delta n_{\rm bio}^{\rm (HL)})/\delta n_{\rm bio}^{\rm (HL)}$, diverges in the limit of large $N$. Such a behavior naturally arises from the $\sqrt{N}$ improvement of the HL over the SIL. Considering Figs.~\ref{fig5}c and d it might seem at first that quantum plasmonic sensing is not necessary because a given resolution can always be achieved by simply increasing the intensity of a classical input source to get the SIL, {\it e.g.} $\delta n_{\rm bio}\sim 10^{-8}$ RIU is in principle achievable by a $\lambda_{0}=810~{\rm nm}$ laser with an initial power of $1$ mW having $N\sim 4\times10^{15}$ photons per second and an appropriate mode volume with a large power density~\cite{Piliarik09,Taylor15,Taylor16}. A high-power source, however, is not commonly desired for biological measurements since it may damage the specimen under investigation~\cite{Neuman99,Peterman03,Taylor15} or cause other unwanted phenomena such as thermal modulation of the surface plasmon mode~\cite{Kaya13}. In this case, at the few-photon level, one may then consider the benefits of using quantum sensing with either a dielectric or plasmonic waveguide. Here, for a fixed value of $N$, the quantum plasmonic sensor provides a low mode volume and allows one to go beyond the SIL in a compact setting below the diffraction limit. This is crucial when only a small biological sample is available, or one would like a more compact and integrated sensing device than standard dielectric components can achieve. Note that while the power density in a plasmonic waveguide is enhanced compared to a dielectric waveguide due to the low mode volume, it is the total power that is important in gaining the quantum advantage, as the shot-noise or Heisenberg limits are related to the photon number statistics rather than the optical power density. Despite all the demanding requirements of quantum measurement with plasmonic systems, in recent years several experimental studies have demonstrated the feasibility of quantum plasmonic sensing using low intensity input sources~\cite{Kalashnikov14,Fan15,Pooser16}.

{\it Discussion---.}
In this work we studied quantum plasmonic sensing in the few photon regime and showed an example quantum plasmonic sensor that can beat the SNL in the presence of metallic losses. We have demonstrated how the inclusion of quantum techniques in a plasmonic system enables one to further improve its sensitivity and resolution beyond the SNL, while keeping the compactness of the device on scales far below the diffraction limit. Our analysis is applicable to any type of plasmonic sensing platform and we leave a variety of technical issues related to quantum plasmonic sensors for future works. For example, the performance of a quantum plasmonic sensor would benefit from further investigation into different excitation platforms such as a prism, grating, localized SPR sensor, metamaterials and graphene, as well as modulation-based approaches. In addition to the sensitivity and resolution considered in this work, other figures of merit such as accuracy, precision, or kinetic analysis will need to be taken into account for practical use in industry. We envisage that progress in quantum metrology will reshape the field of plasmonic biosensing -- a field that has already developed into mature technology for at least two decades~\cite{Taylor16}. Our work opens up a path between the quantum metrology and plasmonic sensing fields, and has the potential to lead to a variety of studies at the level of practical realization. Integrated quantum plasmonic sensors may also find application in on-chip nanoscale quantum network devices, with potential uses in quantum tasks where precise and compact measurement is required~\cite{Gisin07, Sangouard11}.

{\it Acknowledgments.}
C. L. thanks S.-Y. Lee and K. H. Seol for discussions. This research was supported by the Ministry of Science, ICT and Future Planning (MISP) Korea, under the Information Technology Research Center (ITRC) support program IITP-2016-R0992-16-1017 supervised by the Institute for Information and Communications Technology Promotion (IITP), the National Research Foundation of Korea (NRF) Grant funded by the Korea government (MSIP; No. 2014R1A2A1A10050117), the South African National Research Foundation, the South African National Institute for Theoretical Physics, the Marie Sklodowska-Curie Early Stage Researcher programme, the Marie Curie Training Network on Frontiers in Quantum Technologies, and the European Office of Aerospace Science and Technology EOARD.

\appendix
\section{Characteristic equations of nanowire waveguides}
The characteristic equations for the lowest order plasmonic TM mode in a metallic nanowire~\cite{Takahara97,Stockman04} and the photonic LP$_{0m}$ modes in a dielectric nanowire~\cite{Saleh07} are respectively given by 
\begin{eqnarray}
\frac{\epsilon_{m}}{k_{m}}\frac{I_{1}(k_{m}r)}{I_{0}(k_{m}r)} +\frac{\epsilon_{\rm clad}}{k_{\rm clad}}\frac{K_{1}(k_{\rm clad}r)}{K_{0}(k_{\rm clad}r)} &=& 0,\nonumber\\
k_{d}r \frac{J_{1}(k_{d}r)}{J_{0}(k_{d}r)} - k_{\rm clad} r \frac{K_{1}(k_{\rm clad} r)}{K_{0}(k_{\rm clad} r)} &=&0,\nonumber
\end{eqnarray}
where $I_{p}(x)$ and $K_{p}(x)$ are the $p^{\rm th}$-order modified Bessel functions, $J_{p}(x)$ is the $p^{\rm th}$-order Bessel function, and $k_{0}$ is the free space wavenumber of the EM wave ($k_{0} = 2\pi /\lambda_{0} =\omega/c$, where $\lambda_{0}$ denotes the free space wavelength). Here, $k_{m}=k_{0}\sqrt{(k/k_{0})^{2}-\epsilon_{m}(\omega)}$, $k_{d}=k_0\sqrt{\epsilon_{\rm d}-(k/k_{0})^{2}}$, and $k_{\rm clad}=k_{0}\sqrt{(k/k_{0})^{2}-\epsilon_{\rm clad}}$, where $\epsilon_{m}(\omega)$, $\epsilon_{d}$ and $\epsilon_{\rm clad}$ denote the dielectric constants for the metallic core, the dielectric core and the dielectric cladding, respectively. In the main text, the cladding takes the form of the biological medium defined by the refractive index $n_{\rm bio}$. The characteristic equations determine the wavenumber $k$ for the plasmonic or photonic mode as a function of the core radius $r$ and the dielectric functions at the considered wavelength. For simplicity we consider a single-mode dielectric fiber, where the core diameter is small enough for only the fundamental $m=1$ mode to exist, {\it i.e.} $\rm LP_{01}$. This is satisfied by the condition $\sqrt{(k_{d}r)^{2}+(k_{\rm clad} r)^{2}}<2.405$, which is met by the parameters used.


\vskip0.2cm
\section{Output states of two-mode interferometer}
The output state arising from a coherent state $\ket{\alpha}$ input to the Mach-Zehnder interferometer (where a phase acquisition occurs on the first arm) can be obtained by
\begin{eqnarray}
\ket{\alpha}\ket{0} \xrightarrow[]{50/50~BS} \ket{\frac{1}{\sqrt{2}}\alpha}\ket{\frac{1}{\sqrt{2}}i\alpha} \xrightarrow[]{e^{\phi\hat{n}_{1}}} \ket{\frac{1}{\sqrt{2}} \alpha e^{i\phi}}\ket{\frac{1}{\sqrt{2}}i\alpha}  \nonumber\\\xrightarrow[]{50/50~BS}  \ket{\frac{1}{2}\alpha(e^{i\phi}-1)}\ket{\frac{1}{2}i\alpha (e^{i\phi}+1)},\nonumber
\end{eqnarray}
where a beam splitter (BS) operation applies the transformation $\hat{a}_{1}^{\dagger} \rightarrow \frac{1}{\sqrt{2}} (\hat{a}_{1}^{\dagger}+i\hat{a}_{2}^{\dagger})$ and $\hat{a}_{2}^{\dagger} \rightarrow \frac{1}{\sqrt{2}} (i\hat{a}_{1}^{\dagger}+\hat{a}_{2}^{\dagger})$.
On the other hand, the output state arising from inputting a NOON state $\frac{1}{\sqrt{2}}(\ket{N0}+\ket{0N})$ into the two-mode interferometer can be written as 
\begin{eqnarray}
\frac{1}{\sqrt{2}}(\ket{N0}+\ket{0N}) \xrightarrow[]{e^{\phi\hat{n}_{1}}} \frac{1}{\sqrt{2}}(e^{iN\phi}\ket{N0}+\ket{0N}).\nonumber
\end{eqnarray}
For the classical and quantum sensing scenarios, the measurements signals can be given as 
\begin{eqnarray}
\langle\hat{M}\rangle &=& \langle\hat{n}_{2}\rangle - \langle\hat{n}_{1}\rangle
= \abs{\alpha}^{2}{\rm cos}\phi (n_{\rm bio}), \nonumber\\
\langle\hat{A}\rangle &=& {\rm cos}(N\phi(n_{\rm bio})).\nonumber
\end{eqnarray}
The estimation resolutions with respect to the effective phase $\phi(n_{\rm bio})$ can be obtained via the linear error propagation method as
 \begin{eqnarray}
\delta \phi(n_{\rm bio})&=&\frac{\big(\langle\hat{M}^{2}\rangle-\langle\hat{M}\rangle^{2}\big)^{1/2}}{\abs{\partial \langle\hat{M}\rangle / \partial \phi(n_{\rm bio})}}=\frac{1}{\abs{\alpha}{\rm sin}\phi(n_{\rm bio})}\nonumber\\
\delta \phi(n_{\rm bio})&=&\frac{\big(\langle\hat{A}^{2}\rangle-\langle\hat{A}\rangle^{2}\big)^{1/2}}{\abs{\partial \langle\hat{A}\rangle / \partial \phi(n_{\rm bio})}}=\frac{1}{N},\nonumber
\end{eqnarray}
where $\langle\hat{M}^{2}\rangle=\abs{\alpha}^{2} +\abs{\alpha}^{4} {\rm cos}^{2}\phi(n_{\rm bio})$ and $\langle\hat{A}^{2}\rangle =1$. 

\vskip0.2cm
\section{Quantum Fisher information}
The quantum Fisher information is defined as the maximum of the classical Fisher information over all possible generalized measurements $\{\hat{E}_{y}\}$, written as 
\begin{eqnarray}
F_{Q}={\rm max}_{\{ \hat{E}_{y}\}}F(X;\{ \hat{E}_{y}\}),\nonumber
\end{eqnarray}
where $F(X;\{ \hat{E}_{y}\})\equiv \int dy \frac{1}{P(y\vert X)}(\partial P(y\vert X)/\partial X)^{2}$ denotes the classical Fisher information, $P(y|X)={\rm Tr}[\rho(X)\hat{E}_{y}]$ represents the probability of obtaining the outcome $y$ conditional on the value of $X$ upon which the initial state $\rho(X)$ depends, and $\{\hat{E}_{y}\}$ satisfies $\int dy \hat{E}_{y}=\openone$. The quantum Fisher information quantifies the amount of information about the parameter $\phi$ that a state contains when the optimal measurement is performed. For pure states $\vert \psi \rangle$, the quantum Fisher information can be simplified to $F_{Q}=4(\langle \psi' \vert \psi' \rangle - \vert \langle \psi' \vert \psi \rangle \vert ^{2})$, where $\vert \psi' \rangle=\frac{\partial \vert \psi \rangle }{\partial \phi}$. For the state with definite photon number $N$ given by $\ket{\psi}_{in}=\sum_{n=0}^{N} c_{n} \ket{n,N-n}$ and a fictitious beamsplitter with a transmitivity $\eta$ modelling loss in one arm of a two-mode interferometer, the quantum Fisher information is given in terms of $\vert \psi_{\rm out}' \rangle$ by~\cite{Dorner09}
\begin{eqnarray}
F_{Q}
=4\Big( \sum_{n=0}^{N}n^{2} x_{n} -\sum_{l=0}^{N} \frac{(\sum_{n=l}^{N} n x_{n}B_{l}^{n})^{2}}{\sum_{n=l}^{N}x_{n}B_{l}^{n}}\Big),\nonumber
\end{eqnarray} 
where $x_{n}=\abs{c_{n}}^{2}$ and $B_{l}^{n}= {{n \choose l}} {{N-n \choose 0}} \eta^{n} (\eta^{-1}-1)^{l}$.

\vskip0.2cm
\section{Wedge waveguide}
The propagation constant and loss for the silver wedge waveguide was obtained by 2D finite element method (FEM) simulation in COMSOL. The material properties of silver were taken from Rakic {\it et al.}~\cite{Rakic98}. High quality wedge waveguides have recently been fabricated by template stripping silver from etched silicon, so the silicon etching angles were taken as the angles for the waveguides, delivering wedge waveguides with a top angle of 70.6 degrees~\cite{Kress15}. The top was rounded with a radius of curvature of $20~{\rm nm}$. For robustness of the simulations, a height of $5~{\rm \mu m}$ was taken, but the optical mode is highly confined at the tip of the waveguide, allowing to make wedges as small as $50~{\rm nm}$ high without appreciable coupling to the base corners.


\end{document}